\documentclass[article]{IEEEtran}
\usepackage{amssymb,amsmath,amsfonts,bm,epsfig,graphicx,theorem,latexsym}
\usepackage{color,subfigure}
\usepackage{times,authblk}
\usepackage{setspace,multicol}%
\usepackage{verbatim}%
\usepackage{enumerate}
\usepackage{eucal}
\usepackage{caption}
\usepackage{stmaryrd,bbm,subfigure,units}
\usepackage{bbm}
\usepackage{bbold}
\usepackage{graphicx,dblfloatfix}
\usepackage[space]{cite}

\usepackage{pgfplotstable}

\allowdisplaybreaks[1]

\newtheorem{theorem}{Theorem}

\newtheorem{definition}{Definition}

\newtheorem{remark}{Remark}

\newcommand{\mbf}[1]{\mathbf{#1}}
\newcommand{\set}[1]{\mathcal{#1}}

\newcommand{\bfV}{\mathbb{V}}

\renewcommand{\Pr}{\mathbb{P}}
\newcommand{\E}{\mathbb{E}}

\newcommand{\M}{\mathsf{M}}

\renewcommand{\d}{\mbf{d}}

\newcommand{\K}{\set{K}}

\newcommand{\mw}[1]{{\color{blue}{#1}}}

\allowdisplaybreaks[4] 
\graphicspath{{.}{./Figures/}} 

\IEEEoverridecommandlockouts
\IEEEoverridecommandlockouts

\title{State-Adaptive Coded Caching for Symmetric Broadcast Channels}
\author{ $^\dag$Shirin Saeedi Bidokhti, $^{\ddag}$Mich\`{e}le Wigger,  $^{\dag*}$Aylin Yener, and $^{\ddag*}$Abbas El Gamal\\
	\small 
	$^{\dag}$ University of Pennsylvania, saeedi@seas.upenn.edu, \quad $^{\dag*}$The Pennsylvania State University, yener@engr.psu.edu\\
	${\ddag}$ LTCI, Telecom ParisTech, Universit\'e Paris-Saclay, 75013 Paris,  France,
	michele.wigger@telecom-paristech.fr\\
	${\ddag*}$Stanford University, abbas@ee.stanford.edu
	
}

\begin{document}

\maketitle

\begin{abstract}
Coded-caching delivery is considered over a symmetric noisy broadcast channel whose state is unknown at the transmitter during the cache placement phase. In particular, the delivery phase is modeled by a state-dependent broadcast channel where the state remains constant over each transmission block and is learned by the transmitter (and the receivers) only at the beginning of each  block. A \emph{state-adaptive coded caching} scheme is proposed that  improves either on rate  or  decoding latency over two baseline schemes that are based on standard coded caching. 
\end{abstract}
\section{Introduction}
Coded caching \cite{MN14} has recently emerged as a means to improve content delivery in multiuser networks. The performance gains offered by coded-caching scale with the number of users and go beyond those so-called local gains stemming from the fact that part of the data is locally stored at the receivers. While earlier works studied network models with noiseless channels  for delivery \cite{MN14},  caching has more recently been studied in noisy channels, including broadcast channels (BCs) that are most related to this work. In particular, \cite{BidokhtiWiggerTimo17,tulino_jccs,AmiriGunduz17, SaeediWiggerYenerARXIV17}  consider static (and known) degraded BCs and propose joint cache-channel coding schemes that improve rate of communication and attain new global caching gains when the users have  unequal channel qualities and the weaker receivers have  larger cache memories (or demand less data). {Time-varying (fading) BCs and the interplay between feedback, channel state information and spatial multiplexing with caching have also been studied  in \cite{ZhangElia17, GhorbelKobayashiYang17, NgoYangKobayashi17,wang_xian_liu_2015,piovanojoudehclerckx-2017}. These works apply separate cache-channel coding architectures; hence,  the performance of communication in the delivery phase is limited by the weakest users}.  By contrast, in this work we illustrate the benefits of joint cache-channel coding schemes for state-dependent BCs even when different users have equal size caches and i.i.d. channel statistics.

In this work, we model the delivery phase as a state-dependent BC in which the state sequence is constant over a coherence block and  changes from block to block in an i.i.d. manner. This channel model subsumes the standard block fading channel model. The transmitter and receivers learn the realization of the state at the {\it beginning of each block}. This can be done using pilot symbols and feedback. For clarity of presentation, we consider state-symmetric BCs in which all users have equal size caches and statistically equivalent channels and we assume that the channel is degraded in each state. Since the state realizations  vary over blocks, a receiver that is strongest in one block can be weakest in the next block. 

We propose a coding scheme for state-dependent BCs termed {\it state-adaptive coded caching} hereafter. The caching phase of our scheme is performed in an uncoded manner, following the original work of \cite{MN14}. Our  delivery scheme applies (i) opportunistic user scheduling across blocks and (ii) generalized coded caching  \cite{SaeediWiggerYenerARXIV17} in each block. Specifically, only the $t+1$ receivers with the best channel conditions are  served in each block, $t$ being the coded caching parameter used in the cache placement \cite{MN14}.  The proposed scheme serves each of the chosen receivers with a transmission rate that is proportional to its channel quality; i.e., each chosen receiver $k$  is served at a rate that approaches $I(X; Y_k |S=s_b)$, where $X, Y_k, S=s_b$ denote the input, output, and state variables in block $b$, respectively. This performance is achieved by  a variation of Tuncel coding \cite{tuncel06} where for each receiver $k$, the transmitter only encodes bits that are stored in the cache memories of all the other receivers in the chosen subset. This implies a state-adaptive virtual cache allocation at the receivers that allocates a larger portion of cache memories for decoding at weaker receivers than at stronger receivers. Note that for the state-symmetric BC considered in this paper, the total rate and the total required cache size at each user are the same on average (in the long run) across all users.

The proposed strategy is compared to two baseline schemes that combine standard coded caching with the opportunistic BC codes \cite{tsebook} in a separate cache-channel coding architecture. The first baseline scheme, which we term \emph{blockwise coded caching},  operates on a per-block basis and is  limited by the worst channel in each block. {A variant of this baseline scheme in which  opportunistic user selection policy is replaced by a threshold-based user selection policy is discussed in \cite{NgoYangKobayashi17}. Our proposed strategy  also operates on a per-block basis but employs a joint cache-channel coding architecture such that the communication  to stronger users is not limited by weaker users. It} therefore achieves higher rates than  blockwise coded caching. The second baseline scheme, which we term \emph{ergodic coded caching}, codes over the entire communication duration,  i.e., over many blocks. This results in symmetric channel conditions for  all the receivers and eliminates the rate-bottleneck issue of weak receivers in a coherence block. It has, however, the drawback that decoding  is performed only at the end of transmission. 
In   state-adaptive coded caching  (as well as in the first baseline scheme),   decoding can be performed after each block so that a part of the message bits can be recovered earlier. This is particularly beneficial in video streaming in which one wishes to start watching a movie as soon as some of the bits are recovered\footnote{The assumption here is that the movie is encoded using multi-description coding and thus the order of the bits is not relevant. Otherwise, it is also possible to prioritize the bits.
}. 
We quantify this notion by a new delay measure termed the \emph{decoding latency factor} that describes the extent to which decoding is performed sequentially. We show a factor of two improvement in decoding latency factor of  the new state-adaptive coded caching scheme over the second baseline scheme.

\section{Problem Definition}

Consider a state-dependent $K$-receiver broadcast channel (BC) with (finite) input, output, and state alphabets $\set{X}, \set{Y},$ and $\set{S}$.
Given the time-$i$ channel input $X_i\in\set{X}$ and  state   $S_i\in\set{S}$,  receiver~$k\in\mathcal{K}:=\{1,\ldots, K\}$'s time-$i$ output $Y_{k,i}\in\set{Y}$ follows the broadcast channel law
\begin{equation}\label{eq:BC}
p_{Y_k|X,S}(y_{k,i}|x_i,s_i).\end{equation}
For simplicity, we consider state-symmetric BCs in which 
for any permutation on users $\nu\colon \K \to \K$  there exists a permutation on states $\pi_\nu\colon \set{S}\to \set{S}$, so that for all $s\in\mathcal{S}$, $x\in\set{X}$,  $y\in\set{Y}$, we have (see \cite[Definition~1]{KimChiaElGamal} for the two-user definition):
\begin{subequations}\label{eq:symBC}
	\begin{align}
	p_S(s)&=p_S(\pi_\nu(s))\\
	p_{Y_k|X,S}(y|x,s)&=p_{Y_{\nu(k)}|X,S}(y|x,\pi_\nu(s)), \quad \forall k\in\K.
	\end{align}
\end{subequations}
Moreover, we assume that the BC is stochastically degraded \cite{elGamalBook} in any  state realization $S=s$.

The state sequence $S_1, S_2, \ldots,$  stays constant over a coherence interval of  $T_s$ channel uses and  then changes in an independent and identically distributed (i.i.d.) manner. I.e., 
\[S_{(b-1)T_s+1}=\ldots=S_{bT_s}=S^\prime_b,\quad b=1,2,\ldots \]
where $S^\prime_1, S^\prime_2,\ldots$ is an  i.i.d. sequence distributed according to a given distribution $p_{S^\prime}(\cdot)$.

The transmitter has access to a database with $D$ independent  messages (files) $W_1,\ldots, W_D$, each consisting of $nR$ i.i.d.  random bits.
Here,  $n$ denotes the blocklength and $R$ the message rate.  Each receiver~$k$ demands exactly one of the messages, which we denote by $W_{d_k}$.
Receiver $k\in\mathcal{K}$ has access to a local cache memory of $n\M$ bits. 

Communication takes place in two phases. The first \emph{cache placement phase}  is assumed to   take place during a period of low network congestion and  is thus assumed error free.  In this phase, the transmitter  stores information about the messages in each of the $K$ receivers' cache memory. So, in receiver~$k$'s cache memory,   it  stores
\begin{equation*}
\bfV_k := g_k(W_1, \ldots, W_D)
\end{equation*}
 for some function
$g_k \colon  \{1,\ldots, 2^{nR} \}^D \to 2^{n \M}$ that is  known to all terminals. The cache content   $\bfV_k$ is known only to the transmitter and receiver~$k$.
During the placement phase, it is unknown which messages are demanded by the users; so  $g_k$ cannot depend on the demands.

The subsequent \emph{delivery phase} takes place during periods of high network congestion and is modeled by the state-dependent BC in \eqref{eq:BC}.  At the beginning of the delivery phase, each receiver demands one of the messages in the library; i.e., receiver $k$ demands message $W_{d_k}$. At this time, the transmitter and all receivers get informed about all receivers' demands,
$\d=(d_1,d_2,\ldots,d_K)$.
The transmitter then computes the  sequence of channel inputs as 
\begin{equation*}
X_i := f^{(i)}(\d,W_1,\ldots,W_D,S^{i}), \quad i\in\{1,\ldots, n\},
\end{equation*}
where
$
f^{(i)} :  
\{1,\ldots, D\}^K \times \{1,\ldots, 2^{nR}\}^D\times \mathcal{S}^i
\to 
\set{X}.$

Decoding is performed  \emph{online}. In particular, we present coding schemes in which  receivers  recover a certain number of  message bits after each coherence interval $T_s$. Let 
\begin{equation}
B= \frac{n}{T_s}
\end{equation}
denote  the number of coherence blocks encountered when communicating over $n$ channel uses. 
The online decoding procedure is described as follows. After each coherence block $b\in\{1,\ldots, B\}$, 
receiver~$k$ recovers $m_{k,b}$ new bits of its desired message $W_{d_k}$ using the decoding operation
\begin{equation*}
\hat{W}_{k,b} := \varphi_{k,b}(\mathbf{d},Y_k^{bT_s}, \bfV_k,S^{bT_s}),
\end{equation*}
where $\varphi_{k,b} : \{1,\ldots, D\}^K \times \set{Y}^{bT_s} \times \set{V}_k\times \set{S}^{bT_s} \to \{1,\ldots, 2^{m_{k,b}}\}.$
The final estimate of the receiver for message $W_{d_k}$ is then composed of the concatenation of all the estimates
$(\hat{W}_{k,1}, \hat{W}_{k,2}, \ldots  )$.

To capture the nature of online  decoding, we study the following {average expected delay per bit}
\begin{align}\label{eq:Lbit}
\bar{L}_{\textnormal{bit}}:= \max_{\mathbf{d}} \frac{1}{K}\sum_{k=1}^K \E \left[  \sum_{b=1}^n m_{k,b} \cdot bT_s \right] \frac{1}{(R-\M/D)  n} ,
\end{align}
where the worst case over all possible demands is considered.
Normalization is by $(R-M/D)n$ because we wish to average only over the number of transmitted bits but not over the bits that are already stored in the cache memory. Expectation  is over the random  state, channel realizations and  messages.

We consider the worst-case error probability over  demands:
\begin{equation*}
P_{\textnormal{e}}^{(n)}:= 
\max_{\d \in \{1,\ldots, D\}^K} \Pr\bigg[\ 
\bigcup_{k = 1}^K
\big\{ \hat{W}_k \neq W_{d_k} \big\}\
\bigg].
\end{equation*}
We also assume that  the coherence time $T_s$ and the number of blocks $B$ 
 tend to infinity, i.e.,
$T_s,B \to \infty.$
Under this assumption, for positive rates $R>0$,  the delay $\bar{L}_{\textnormal{bit}}$ also  tends to infinity. We, therefore, further normalize it  by the blocklength $n$, yielding  the \emph{decoding latency factor $\bar{\rho}$:} 
\begin{IEEEeqnarray}{rCl}
	\bar{\rho}  \triangleq \lim_{n\to \infty} \frac{\bar{L}_{\textnormal{bit}} }{n}.
\end{IEEEeqnarray}

\begin{definition}
A triple $(\M, R, \rho)$ is  achievable, if there exists a sequence (in $n$) of caching and delivery encoders  and decoders with cache and message rates $\M$ and $R$ such that  
\begin{align}
\lim_{n\to \infty }P^{(n)}_{\textnormal{e}} =0 \qquad \textnormal{and}\qquad 
\bar{\rho}\leq \rho.
\end{align}

\end{definition}

\section{State-Adaptive  Coded-Caching} \label{sec:scheme}
Our proposed scheme has a parameter $t\in\{0, \ldots, K-1\}$, where $t+1$ indicates the number of users that are simultaneously served in the delivery phase. E.g., parameter $t=0$ corresponds to opportunistic broadcasting, which is known to achieve the maximum sum-rate and symmetric rate\cite[Chapter~6]{tsebook} when there are no cache memories.

We start with some definitions. Fix $t\in\{0,\ldots, K-1\}$.
Let 
$	\set{G}_1^{ t}, \ldots, \set{G}_{ { K \choose  t}}^{ t}$
	denote all  size-$t$ subsets of~$\K$, i.e., all sets of $t$ users. 
Choose a conditional probability law $p_{X|S}$ so that
\begin{equation}
p_{X|S}( x|s)=p_{X|S}(x|\pi_\nu(s)),\qquad  \forall x\in \mathcal{X}, \ s\in\mathcal{S},\label{sym-input}
\end{equation}
{for any set of permutations $\pi_\nu$  introduced in \eqref{eq:symBC}.} 

Define the mapping $G^{t+1}\colon \set{S} \to \K^{t+1}$ such that for all $s\in\set{S}$, $k \in G^{t+1}(s)$, and $j \in (\set{K}\backslash G^{t+1}(s))$, the channel $p_{Y_j|X,S=s}$ is stochastically degraded with respect to $p_{Y_k|X,S=s}$. 
In our scheme,  $G^{t+1}(s)$ is the set of ``active" receivers  in a block with state $
S'=s$.  The symmetry condition \eqref{eq:symBC} ensures that $G(S)$ is uniformly distributed over all $t+1$-user sets of $\K$.

Define for  $k\in\K$:
\begin{align}\label{eq:ratea}
R= \frac{t+1}{K-t} \cdot I\big(X;Y_k|S, \{ k\in G^{t+1}(S)\}\big) -\epsilon.
\end{align}
Note by \eqref{eq:symBC} and \eqref{sym-input} that  the choice of $k$ does not matter. Here, $ \{ k\in G^{t+1}(S)\}$ denotes the event that index $k$ is an element of $G^{t+1}(s)$.  Let the cache size be
\begin{align}\label{eq:memorya}
\M = \frac{{{K-1}\choose t-1}}{{K\choose t}}RD=\frac{t}{K}RD.
\end{align}

Distribute the $nR$ bits of file  $W_d$  into ${K \choose t}$ queues
$Q_{d,\set{G}_1^{t}}, \ldots, Q_{d,\set{G}_{{K\choose t}}^{t }},$
each consisting of ${nR}\cdot{{K \choose t}}^{-1}$ bits\footnote{{We assume $n\geq {K\choose t}$ since our interest is in the regime $n\to \infty$. }}

\noindent\underline{\textit{Placement Phase:}} For each $k$ and  $\ell \in \big\{1,\ldots, {K\choose t}\big\}$ such that $k \in \mathcal{G}^{t}_{\ell}$, store all the bits of queue $Q_{d,\set{G}_\ell^{t}}$ in receiver~$k$'s cache memory. The cache content of user $k$ is thus: 
\begin{IEEEeqnarray}{rCl}\label{eq:cached}
\lefteqn{\bfV_{k} = \Big\{Q_{d,\set{G}_\ell^{t}}\colon d\in\{1,\ldots, D\} \textnormal{ and }}\qquad \nonumber \\
 &  & \hspace{1.8cm}\ell \in \bigg\{1,\ldots, {K\choose t}\bigg\} \; \textnormal{ s.t. }\; k \notin \set{G}_{\ell}^{t}\bigg\}.\IEEEeqnarraynumspace
\end{IEEEeqnarray}
Notice that each sub-message is stored at exactly $t$ receivers. Moreover,  the placement of the information  does not depend on the realization of the channel state. By \eqref{eq:memorya}, this placement satisfies the memory constraint of $n\M$ bits.

\noindent\underline{\textit{Delivery Phase:}}
Delivery is block-by-block in our scheme. 
Consider the coherence block $b\in\{1,\ldots,B\}$ and 
assume that the channel state is realized to be $S_b'=s_b$. At the beginning of each coherence block, the transmitter retrieves  the next
\begin{IEEEeqnarray}{rCl}\label{eq:achM}
	&	\mu_{k,G^{t+1}(s_b)}\triangleq T_s \cdot\left( I\big(X;Y_{k}|S=s_b\big) -\epsilon \cdot \frac{K-t}{t+1} \right)
	\label{eq:Mach}
\end{IEEEeqnarray}
bits from queue $Q_{d_k, G^{t+1}(s_b)\backslash \{k\}}$, for  $k\in G^{t+1}(s_b)$.  
 Denote the  bits retrieved from queue $Q_{d_k, G^{t+1}(s_b)\backslash \{k\}}$ by $W_{k,b}$.  
If  the queue is empty, let  $W_{k,b}$ be the all-zero string.
%
%

Use a random codebook 
\begin{equation}
\set{C}_b^{T_s} = \Big\{  \mathbf{x}_b^{T_s}( w) \colon  \ w \in \big\{1,\ldots,2^{T_s r(s_b)}\big\}\Big\}
\end{equation}
of rate 
\begin{equation}
r(s_b)\triangleq 
\displaystyle \max_{k\in G^{t+1}(s_b)}I(X;Y_k|S=s_b)-\epsilon\cdot \frac{K-t}{t+1}
\end{equation}
with entries  drawn i.i.d. according to a given law $p_{X|S}(\cdot|s_b)$. 
The transmitter  sends the codeword 
\begin{equation}
 \mathbf{x}_b^{T_s}\Big(  \displaystyle{\overline{\bigoplus}_{k \in G^{t+1}(s_b)}}  {W}_{k,b} \Big)
\end{equation}
over the channel.
Here ${\overline{\bigoplus}}$ describes the XOR operation of the submessages after zero-padding to the same length.

Decoding is done sequentially after each block $b=1,2,\ldots$. Consider decoding at receiver~$k\in\K$. Suppose $S_b'=s_b$ and  $k\in G^{t+1}(s_b)$.  Receiver $k$ can  retrieve  bits from the queues 
\begin{IEEEeqnarray}{rCl}\label{eq:cachedk}
	\bigg\{  Q_{d_k, \set{G}_{\ell}^{t}} \colon \;  \; \ell \in \bigg\{1,\ldots, {K \choose t}\bigg\} \;\textnormal{ s.t. } \; k\in  \set{G}_\ell^{{t}}\bigg\} \IEEEeqnarraynumspace
\end{IEEEeqnarray}
that are stored in its local cache.
To recover the missing bits, it uses the retrieved bits to form  the XOR-message
 \begin{equation}\label{eq:retrieve}
 W_{\textnormal{XOR},b}(k):=	\displaystyle  \overline{ \bigoplus}_{{i \in{G}^{t+1}(s_b) \backslash \{k\}} } \;
 W_{i, b}.
 \end{equation}
It then extracts a subcodebook $\tilde{\set{C}}_{b,k}\big( W_{\textnormal{XOR},b}(k)\big)$ from $\set{C}_b$ that contains  all codewords that are compatible with  $W_{\textnormal{XOR}, b}(k)$: 	
 \begin{IEEEeqnarray*}{rCl}
 	\tilde{\set{C}}_{b,k}\big( W_{\textnormal{XOR}, b}(k)\big) &:=& \Big\{ \mathbf{x}_b^{T_s}\big( w \;  \bar{\oplus} \; W_{\textnormal{XOR}, b}(k)  \big) \Big\}.
 \end{IEEEeqnarray*}
 Finally, it  collects the outputs in coherence block $b$, 
{and applies a  maximum likelihood decoder based on  the extracted subcodebook $\tilde{\set{C}}_{b,k}\big( W_{\textnormal{XOR}, b}(k)\big)$  to recover the bits $W_{k, b}$.} If $k \notin G^{t+1}(s_b)$,  receiver~$k$ does not decode anything in this block~$b$. 

\textit{Performance Analysis:} Given that $S_b'=s_b$, the number of bits $m_{k,b}$ recovered by a given receiver~$k\in\K$  at the end of  coherence block $b$ is
\begin{IEEEeqnarray}{rCl}
	m_{k,b} & = & \begin{cases} 0, &\textnormal{ if } k \notin G^{t+1}(s_b) \\ T_s I\big(X;Y_{k}|S=s_b\big)-\frac{\epsilon T_s (K-t)}{t+1} & \textnormal{ if } k \in G^{t+1}(s_b). 
	\end{cases}\nonumber 
\end{IEEEeqnarray}
These bits pertain to queue $Q_{d_k,G^{t+1}(s_b)\backslash\{k\}}$ and are useful information bits unless this queue is empty. 
	
	 Notice that the symmetry conditions \eqref{eq:symBC} and \eqref{sym-input} imply that $\sum_{s \colon k \in G^{t+1}(s)}  p_S(s) I(X;Y_k|S=s)$ does not depend on the receiver index $k$. Moreover, \eqref{eq:symBC} ensures  that the set $G^{t+1}(S)$ is uniformly distributed over all  $t+1$-user subsets  of $\K$. As a consequence, when averaged over the random state realization, for each block $b$  the same expected number of bits is transmitted from each of the queues  $\big\{Q_{d_k, \mathcal{G}_{\ell}^t}\colon k\notin  \mathcal{G}_{\ell}^t\big\}$.
By the ergodicity of the process $\{S_b'\}$ and because during the initialization procedure each queue is allocated the same number of bits, when $B \to \infty$, almost all transmitted bits are useful information bits and all queues will be emptied at the end of the transmission as long as the message rate $R$ satisfies:
\begin{IEEEeqnarray}{rCl}
R&<&\lefteqn{\varlimsup_{B\to \infty} \frac{1}{B T_s} \sum_{b=1}^B m_{k,b} +\frac{\M}{D}}\quad \nonumber \\
&=& \!\!\!\!\sum_{s \in \set{S} \colon k\in G^{t+1}(s)} \!\! p_S(s)\bigg(I\big(X;Y_{k}|S=s\big)-\frac{\!\epsilon (K\!-\!t)\!}{t+1} \bigg)+\frac{\M}{D}\nonumber \\
&\stackrel{(a)}{=}&I(X;Y_k|S, \{ k\in G^{t+1}(S)\}) \Pr[k \in G^{t+1}(S)]\nonumber \\[1.1ex]
 && -\frac{\!\epsilon (K\!-\!t)}{t+1} \Pr[k \in G^{t+1}(S)]+ \frac{\M}{D} \nonumber \\
&\stackrel{(b)}{=}&\frac{t+1}{K} I(X;Y_k|S, \{ k\in G^{t+1}(S)\}) +\frac{\M}{D}- \frac{K-t}{K} \epsilon\nonumber \\
&\stackrel{(c)}{=}&\frac{t+1}{K-t} I(X;Y_k|S, \{ k\in G^{t+1}(S)\}) -\epsilon,
\IEEEeqnarraynumspace
\end{IEEEeqnarray}
where  $(a)$ holds because $\Pr[S=s, k \in G^{t+1}(s)] = \Pr[S=s]$ and $I(X;Y_k|S=s, \{k\in G^{t+1}(s)\})=I(X;Y_k|S=s)$  for all $s$ such that $k\in G^{t+1}(s)$; $(b)$ holds because $\Pr[k \in G^{t+1}(S)]  = \frac{t+1}{K}$ and $(c)$ holds by \eqref{eq:memorya}.

Notice further that by the choice in \eqref{eq:achM}, the probability of decoding error in each block tends to 0 as $T_s \to \infty$.

For finite $n=T_s B$ the   average expected delay per bit is: 
\begin{IEEEeqnarray}{rCl}
\bar{L}_{\textnormal{bit}}
&\stackrel{(a)}{=} & \frac{1}{K}\sum_{k=1}^K \frac{1}{(R-\M/D) n}   \sum_{b=1}^B    \E_{S_b}[ m_{k,b} ] \cdot b T_s   \nonumber \\
& \stackrel{(b)}{=} & \frac{1}{B T_s } \cdot  T_s^2  \cdot \sum_{b=1}^{B} b= \frac{T_s (B+1)}{2 },
\end{IEEEeqnarray}
where $(a)$ follows by the definition in \eqref{eq:Lbit}; and $(b)$ because $  \E_{S_b}[ m_{k,b} ]= R-\M/D$ and $n=B T_s$. 
The decoding latency factor of the proposed	state-adaptive coded caching is thus:
	\begin{IEEEeqnarray}{rCl}
		\bar{\rho}&=& \lim_{T_s,B \to \infty} \frac{(B+1)T_s}{2n}
=\lim_{T_s,B\to \infty} \frac{(B+1)T_s}{2 B T_s} 
	=	 \frac{1}{2} . \IEEEeqnarraynumspace
	\end{IEEEeqnarray} 
	
For each $t\in\{0,1,\ldots, K-1\}$, define
		\begin{IEEEeqnarray}{rCl}
			R_t&:=& \frac{t+1}{K-t} \cdot \max_{\substack{p_{X|S} \colon\!\! \!\phantom{(}\eqref{sym-input} \textnormal{ holds}}} I\big(X;Y_1|S, \{ 1 \in G^{t+1}(S)\}\big)\label{eq:Rthm}\IEEEeqnarraynumspace \\
			\M_t &: = & \frac{t}{K} D R_t\label{eq:Mthm}
			\end{IEEEeqnarray}

			By time/memory-sharing arguments \cite{MN14} and by optimizing over $p_{X|S}$, the presented analysis (with $\epsilon \to 0$) establishes the following theorem.
	\begin{theorem}\label{thm1}

	\label{thm}
		 State-adaptive coded caching  achieves all rate-memory pairs on the upper convex envelope of 
		\begin{IEEEeqnarray}{rCl}
			\{(R_t, \M_t) \colon t =0, 1, \ldots, K-1\}
			\end{IEEEeqnarray}
			with decoding latency factor $\bar{\rho} =  \frac{1}{2}$.
	\end{theorem}
	
\begin{remark}
Including input distributions $p_{X|S}$ that satisfy \eqref{sym-input} but  don't maximize \eqref{eq:Rthm} does not increase the set of achievable rate-memory pairs in Theorem~\ref{thm1}, because they are  subsumed by the upper convex envelope operation.
\end{remark}
\begin{remark}
	The maximization  in \eqref{eq:Rthm} can be re-written as: 
	 \begin{align}
	 \max_{p_{X|S}}\frac{1}{K}\sum_{k=1}^K I\big(X;Y_k|S, \{ k \in G^{t+1}(S)\}\big)\label{eq:equivalentrate}
	 \end{align} 
	 where the maximization is over all (also non-symmetric) input distributions. This follows from the symmetry condition (2). 
	\end{remark}

\begin{remark} The proposed scheme only serves the best $t+1$ receivers in each block. We could combine transmissions to various sets of $t+1$ receivers in a single block by means of superposition or Marton coding. But since for each state realization the BC is assumed degraded, these techniques  do not increase the set of achievable rate-memory-latency triples.

\end{remark}
\section{Comparison to Baseline Schemes}
	Two  baseline schemes derived from standard coded caching are described and compared to the proposed state-adaptive coded caching scheme. The results are summarized in Table~\ref{tab:comparison}.

\subsection{Blockwise Coded Caching}
\label{baseline1}
Fix a parameter $t\in\{0,\ldots, K-1\}$. Consider a separate cache-channel coding scheme with placement strategy as  in Section \ref{sec:scheme} and a delivery strategy that combines standard coded caching  \cite{MN14} of parameter $t$ with  an opportunistic BC code  that in each coherence block serves only the $t+1$ strongest receivers. Specifically, it sends an XOR-message produced by the coded caching algorithm to these strongest $t+1$ receivers.
With this scheme, the performance in each block is  limited by the worst of the $t+1$ best receivers.  In fact, at the end of  coherence block $b$ with state $S_b'=s_b$, the number of bits recovered at receiver $k\in G^{t+1}(s_b)$ is:
\begin{IEEEeqnarray}{rCl}
	m_{k,b}=  \begin{cases} 0, &\!\!\!\! \textnormal{ if } k \notin G^{t+1}(s_b) \\ T_s {\displaystyle\min_{ j\in G^{t+1} (s_b)}  I\big(X;Y_{j}|S=s_b\big)} \nonumber \\[1.4ex] -\frac{\!\epsilon T_s (K-t)}{t+1} &  \textnormal{ if } k \in G^{t+1}(s_b). 
	\end{cases}\nonumber 
\end{IEEEeqnarray}
By symmetry, and when $B\to \infty$, for any $k\in\K$ the message rate to receiver $k$ is:
\begin{align}
R =&\!\!\! \sum_{s \in \mathcal{S} \colon 1\in G^{t+1}(s)}\!\!\!\! p_S(s)\bigg(\!
 \min_{j \in G^{t+1}(s_b)}\! I(X;Y_j |S\!=\!s) -\frac{\!\epsilon(K\!-\!t)\!}{t+1)}\bigg)  \nonumber \\
 & \!+\!\frac{\M}{D}. \label{eq:rateb}
\end{align}

The required cache size $\M$ and the decoding latency factor  are similar as  for the state-adaptive 
coded caching scheme:
\begin{IEEEeqnarray}{rCl}
	\M& =&  \frac{t}{K} R D \qquad \textnormal{and} \qquad 	\bar{\rho}= \frac{1}{2}.\label{eq:memoryb}
	\end{IEEEeqnarray}
		Plugging \eqref{eq:memoryb} into \eqref{eq:rateb}, taking $\epsilon \to 0$, and optimizing over $p_{X|S}$ yields the desired value for the rate $R_t$ in Table~\ref{tab:comparison}.

\subsection{Ergodic Coded Caching}

Fix a parameter $t\in\{0,\ldots, K-1\}$. The scheme combines standard coded caching  with an opportunistic BC code that codes over the entire blocklength $n$. That means,  in each block transmission is only to the best $t+1$ receivers, but decoding is performed only at the end of the entire blocklength $n$. That means, the XOR-message sent to a given a set of $t+1$ receivers is decoded based on \emph{all the blocks} where the opportunistic scheduling chooses to transmit to these $t+1$ receivers. This allows to exploit the ergodic behaviour of the blocks. 
Ergodic coded caching achieves the same rate-memory pairs as state-adaptive coded caching. 
The price to pay   is the worst case decoding latency factor $\bar{\rho}=1$.

\mw{\begin{table*}[h!]
\centering
\begin{tabular}{c | c | c}
	Scheme & Expected Rate $R_t$ & Decoding Latency Factor $\rho$\\ \hline \hline 
	State-Adaptive Coded Caching &  $\phantom{\Bigg(}\displaystyle \frac{t+1}{K-t} \cdot \max_{\substack{p_{X|S} \colon\!\! \!\phantom{(}\eqref{sym-input} \textnormal{ holds}}} I(X;Y_1|S, \{ 1 \in G^{t+1}(S)\})$&  1/2\\ \hline 
	Blockwise Coded Caching & $ \displaystyle \phantom{\Bigg(} 
 \sum_{s \in \set{S} \colon 1\in G^{t+1}(s)} p_S(s)   \;\max_{\substack{p_{X|S} \colon\!\! \!\phantom{(}\eqref{sym-input} \textnormal{ holds}}}	\quad \; \min_{j \in G^{t+1}(s)}  I(X;Y_j |S=s )$ &1/2 \\ \hline 
		Ergodic Coded Caching&$\phantom{\Bigg(}\displaystyle \frac{t+1}{K-t} \cdot \max_{\substack{p_{X|S} \colon\!\! \!\phantom{(}\eqref{sym-input} \textnormal{ holds}}} I(X;Y_1|S, \{ 1 \in G^{t+1}(S)\})$& 1\\  \hline 
\end{tabular}
\caption{Comparison of rate and decoding latency factor for the different coded-caching adaptations.}
\label{tab:comparison}
\vspace{1mm}

\hrule

\end{table*}
}

\section{Gaussian Fading Channels}\label{sec:fading}
	Consider a Rayleigh block-fading channel
	\begin{align}
	Y_{k,i}=h_{k,i}X_{i}+Z_{k,i},\label{fading}
	\end{align}
	with channel coefficients that remain constant over a block,
	\begin{align}
	h_{k,i}=h^\prime_{k,b} ,\qquad \forall i=(b-1)T_s+1,\ldots,bT_s,
	\end{align}
and with $\{h^\prime_{k,b}\}$  an i.i.d.  complex Gaussian sequence with zero-mean unit-variance symbols. The noise sequence $\{Z_{k,i}\}$ is also i.i.d. complex Gaussian of unit variance. Inputs $X_1,\ldots, X_n$ are subject to an expected average block power constraint $P$. 
	
	Let $\mathbf{h}^\prime:=(h^\prime_1,\ldots,h^\prime_k)$ and fix  $t\in\{0,\ldots, K-1\}$.  Here, $G^{t+1}(\mathbf{h}^{\prime})$ denotes the  set of $t+1$ users with largest channel coefficients in $\mathbf{h}^{\prime}$.
		The maximum in Theorem~\ref{thm} is attained by a zero-mean Gaussian input of state-dependent instantaneous power $P(\mathbf{h}^\prime)$, which  can be found  using  the Karush-Kuhn-Tucker conditions  on the equivalent maximization problem \eqref{eq:equivalentrate}. This proves achievability of the upper convex envelope of all  rate-memory pairs 
	\begin{align}
	&R_t=\frac{t+1}{K-t}\mathbb{E}_{\mathbf{h}^\prime}\left[\log\left(1+|h^\prime_1|^2P(\mathbf{h}^\prime)\right)\big| \{1 \in G^{t+1}(\mathbf{h}^{\prime})\}\right]\label{eq:fadingrate}\\
	&M_t=\frac{t}{K}DR.
	\end{align}
where  $P(\mathbf{h}^\prime)$ is the waterfilling solution characterized by:
 \begin{align}
		&\lambda=\sum_{k\in G^{t+1}(\mathbf{h}^{\prime})}\frac{1}{x(\mathbf{h}^\prime)+\frac{1}{|h^\prime_k|^2}}\\
	&P(\mathbf{h}^\prime)=[x(\mathbf{h}^\prime)]^+\label{eq:P1}\\
	&P=\mathbb{E}_{\mathbf{h}^\prime}\left[P(\mathbf{h}^\prime)\right].\label{eq:P2}
		\end{align}

	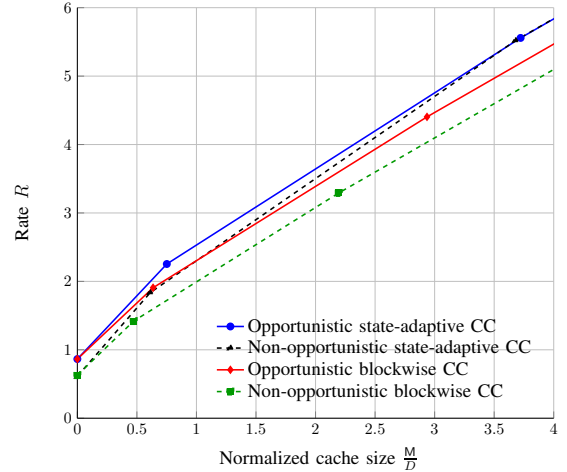
\begin{figure}[t!]
	\centering
	\begin{tikzpicture} [every pin/.style={fill=white},scale=0.6]
  \begin{axis}[scale=.97,
width=0.6\textwidth,
scale only axis,
xmin=0,
xmax=4,
xmajorgrids,
xlabel={\large{ Normalized cache size $\frac{\M}{D}$}},
ymin=0,
ymax=6,
ymajorgrids,
ylabel={\large{Rate $R$}},
axis x line*=bottom,
axis y line*=left,
legend pos=south east,
legend style={draw=none,fill=none,legend cell align=left, font=\large}
]

         \addplot[color=blue,solid,line width=1pt,mark=*]
 table[row sep=crcr]{  
      0 0.8648\\
      .7512 2.254\\
 3.722 5.558\\
 4.722 6.558\\
               };
    \addlegendentry{Opportunistic state-adaptive CC}
%

    %
     \addplot[color=black,dashed,line width=1pt,mark=triangle*]
 table[row sep=crcr]{  
      0 0.6131\\
      .6118 1.835\\
 3.679 5.519\\
 4.679 6.519\\
               };
    \addlegendentry{Non-opportunistic state-adaptive CC}


 \addplot[color=red,solid,line width=1pt,mark=diamond*]
 table[row sep=crcr]{
           0 0.8648\\
 0.6358 1.907\\
     2.936 4.404\\
     4.936 6.404\\
               };
                   \addlegendentry{Opportunistic blockwise CC}

 \addplot[color=black!40!green,dashed,line width=1pt,mark=square*]
 table[row sep=crcr]{
          0 0.63\\
0.4732 1.42\\
     2.195 3.292\\
     4.195 5.292\\
               };
    \addlegendentry{Non-opportunistic blockwise CC}
%
%

\end{axis}

\end{tikzpicture}
	\caption{Comparison of achievable rates on a 3-user example with power constraint $P=4$.\vspace{-.4cm}}
	\label{fig:plot}
	\end{figure}

	Figure~\ref{fig:plot} compares the rates achieved by  state-adaptive and blockwise coded caching (CC) {under  opportunistic and non-opportunistic designs}.  A non-opportunistic design refers to a variation of the schemes where time-sharing is applied in each block  to serve  all subsets of $t+1$ users during the same fraction of   time.
	Each marked memory-rate point corresponds to a choice of the parameter $t$, with the left-most point corresponding to $t=0$ and the right-most point corresponding to $t=K-1$. The curve is obtained by time/memory-sharing between the points. The rate-memory pairs lying to the right of the right-most ($t=K-1$) point are achieved by a scheme that stores a part of each message in  every cache memory and applies placement and delivery strategies  with parameter $t=K-1$ to  the remaining part of the files.

\section*{Acknowledgment}
M. Wigger was supported by the ERC-project CTO-Com.

\bibliographystyle{IEEEtran}

\end{document}